\documentclass[%
 aip,
 apl,%
 amsmath,amssymb,
 reprint,%
]{revtex4-1}

\usepackage{graphicx}% Include figure files
\usepackage{dcolumn}% Align table columns on decimal point
\usepackage{bm}% bold math
\newcommand{\degree}{$^{\circ}$}

\begin{document}

\preprint{AIP/123-QED}

\title{Reply to "Comment on the 'Decrease of the surface resistance in superconducting niobium resonator cavities by the microwave field'"}

\author{G. Ciovati}
\email{gciovati@jlab.org}
\affiliation{Thomas Jefferson National Accelerator Facility, Newport News, Virginia 23606, USA}
\author{P. Dhakal}
\affiliation{Thomas Jefferson National Accelerator Facility, Newport News, Virginia 23606, USA}
\author{A. Gurevich}
\affiliation{Department of Physics, and Center for Accelerator Science, Old Dominion University, Norfolk, Virginia 23529, USA}

\date{\today}% It is always \today, today,

\begin{abstract}

In a recent comment Romanenko and Grassellino  \cite{rom_com} made unsubstantiated statements about our work [Appl. Phys. Lett. \textbf{104}, 092601
(2014)] and ascribed to us wrong points which we had not made. Here we show that the claims of Romanenko and Grassellino are based on 
misinterpretation of Ref. \onlinecite{apl_Qrise}, and inadequate data analysis in their earlier work  \cite{rom}.   

\end{abstract}

\maketitle

The goal of Ref. \onlinecite{apl_Qrise} was to reveal mechanisms of the microwave enhancement of the quality factor $Q(H)$ observed on Ti-alloyed Nb cavities. This was done using the standard Arrhenius method to deconvolute a thermally-activated and residual contributions to the surface resistance \cite{apl_Qrise},
\begin{equation}
R_s(T_s)=Ae^{-U/kT_s}+R_i.
\label{eq:one}
\end{equation}
The Arrhenius method was used in Ref. \onlinecite{gigi} to separate the temperature-independent residual resistance $R_i$ from the conventional BCS contribution $R_{BCS}(T)\simeq Ae^{-U/kT_s}$ measured in Nb cavities at low $H =4$~mT. This procedure was based on the Mattis-Bardeen expression \cite{mb} for $R_{BCS}(T)$, which enables one to unambiguously separate the quasiparticle contribution $R_{BCS}$ from additional temperature-independent contributions to $R_i$ at $T\ll T_c$ which are not described by the simplest version of the BCS model. In Refs. \onlinecite{rom,apl_Qrise}  the Arrhenius method was adopted to analyze $R_s(T, H)$ at strong rf fields in the region of nonlinear electromagnetic response. In this case no simple theoretical expression for the quasiparticle contribution is available so the physical meaning of the phenomenological parameters in Eq. (\ref{eq:one}) becomes far from obvious. This is because the thermally-activated contribution $A(H)\exp[-U(H)/kT_s(H)]$ becomes highly nonlinear in $H$ and gets intertwined with complex nonequilibrium kinetics of quasiparticles \cite{kopnin}, and with such extrinsic mechanisms as trapped vortices, proximity coupled normal oxide regions, etc. One of the manifestations of nonequilibrium effects is rf heating which makes the local temperature of quasiparticles $T_s$ higher than the bath temperature $T_0$, the electron temperature can be higher than the lattice temperature at $kT<<U$ ( Ref. \onlinecite{kopnin}). 

In our work we stated two facts: 1. Alloying Nb cavities with Ti or N significantly extends the field region where $Q(H)$ exhibits a remarkable increase with $H$, 2. Heating effects were disregarded in the analysis of Ref. \onlinecite{rom}.  Contrary to the assertion of Romanenko and Grassellino, we did not make any comments on the validity of Ref. \onlinecite{rom}, nor did we suggest that the mechanism of the significant increase of $Q(H)$ observed on Ti and N alloyed cavities \cite{Dhakal,Anna} is the same as the low-field increase of $Q(H)$ observed on non-alloyed Nb cavities \cite{Padamsee2009}. And we certainly did not suggest that the microwave suppression of $R_s$ is due to heating. What we did say was that it is important to separate intrinsic mechanisms of rf nonlinearity from heating effects to reveal the physics of the microwave suppression of $R_s(H)$.  In any case, the Arrhenius method in which heating was taken into account\cite{apl_Qrise} cannot be less accurate than the same method in which heating was disregarded\cite{rom}, and the claim of Romanenko and Grassellino that taking heating into account can somehow produce some unspecified "systematic errors" was not substantiated.  

The reason why our measurements of $Q(H)$ were performed in the extended temperature region 1.6~K $ < T < 5$~K with 20-30 $T$-datapoints per each rf field value, and a self-consistent account of rf heating is that the Arrhenius fit becomes far more reliable than what was done in Ref. \onlinecite{rom} where heating was disregarded and data were taken only in a very narrow temperature range 1.6~K $<T<2$~K per each rf field value \cite{rom}. The actual number of $T$-datapoints per each rf field value was not specified: the inset in Fig. 1b of Ref. \onlinecite{rom} shows $R_{BCS}(H,T)$ for only four T-datapoints at 1.7, 1.8, 1.9 and 2 K, which would be insufficient for a stable fit. In any case, the accuracy of $U$ extracted from the Arrhenius fit of $\ln R_s(T)$ in such narrow temperature region, $1.6<T<2$ K is poor, but the accuracy of evaluation of $A$ and $R_i$ in Eq. (\ref{eq:one}) from the extrapolation of the semi-logarithmic plot of $\ln R_s$ versus $1/T$ is much worse. 

The analysis of our own data shows that, had we restricted our measurements to 1.6~K~$<T<2$~K like in Ref. \onlinecite{rom}, the accuracy and the scattering of the fit parameters $A$ and $R_i$ with nine T-datapoints per each rf field value would have been so bad that no reliable conclusions about the physical mechanisms of $R_s(H)$ could have been made, even though Eq. (\ref{eq:one}) still provides a good fit to the $R_s(T)$ data, as illustrated in Figs. 1 and 2. This is because there are not enough datapoints to constrain the fit parameters for the acceptable chi-square statistical minimization for which it is important to have enough data points above $T_{\lambda}=2.17$~K, where the exponential temperature dependence of $R_s(T)$ dominates and rf heating must be taken into account. Otherwise, the fit parameters $A$ and $R_i$ become poorly constrained, leading to erroneous dependencies of $A$, $U$ and $R_i$ on $H$. This issue, combined with the increased systematic uncertainty in measuring quality factor value of $\sim 10^{11}$ at $\sim 1.6$~K,  shows that the physical conclusions inferred from the $R_s$-decomposition of Ref.~\onlinecite{rom} can hardly be trusted. This problem with the procedure of Ref. \onlinecite{rom} may explain the negative residual resistance resulting from a fit of one of the data set shown in Fig. 6 of Ref. \onlinecite{rom_bake}, the unrealistically large ratio $\Delta/kT_c = 2.4$ for Nb at 20 mT and non-systematic oscillations of $U(H)$ and $R_i(H)$ shown in Fig. 1a of Ref. \onlinecite{rom}.

\begin{figure}
\includegraphics[width=3.3in]{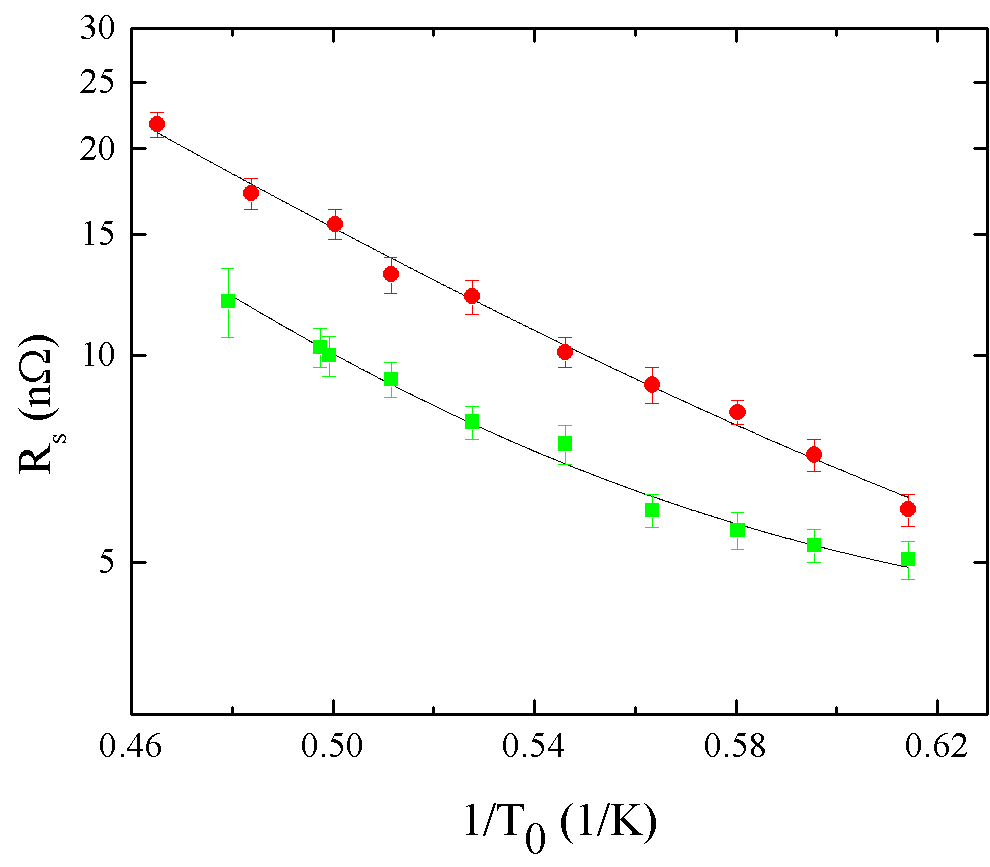}
\caption{\label{fig:one} Subset of mean $R_s(T_0)$ data below $T_{\lambda}$ after 1400~\degree C HT at $\mu_0H=6.2 \pm 0.4$~mT (circles) and $\mu_0H=27 \pm 1$~mT (squares) \cite{apl_Qrise}. Solid lines were obtained from a fit with Eq.~(\ref{eq:one}) using $T_s=T_0$.}
\end{figure} 
\begin{figure}
\includegraphics[width=3.3in]{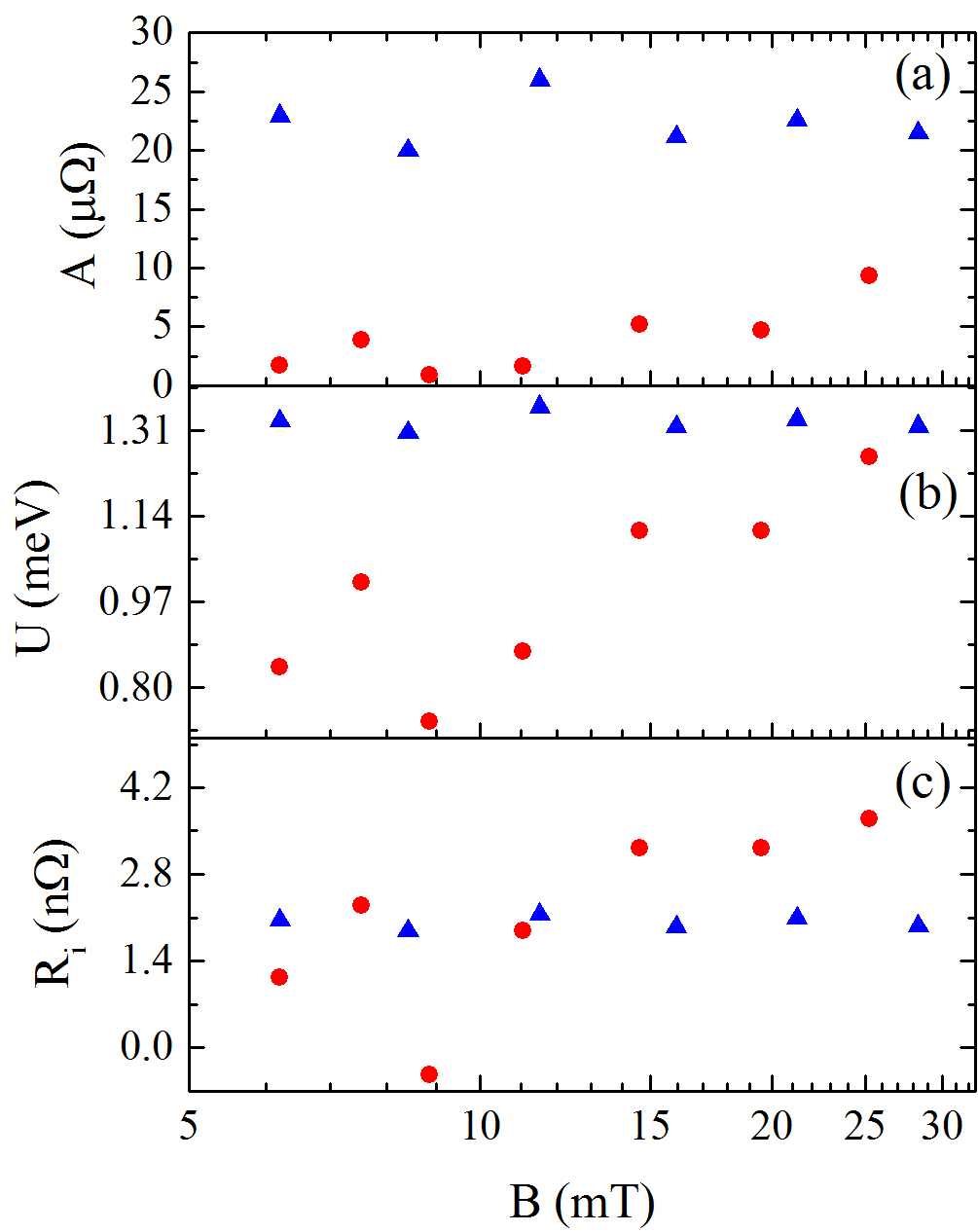}
\caption{\label{fig:two} Dependencies of $A$, $U$ and $R_i$ on the rf field amplitude after 1400~\degree C HT (circles) and after $\sim$1~$\mu$m BCP (triangles) obtained by fitting  only the data below $T_{\lambda}$ with Eq.~(\ref{eq:one}), like in Refs.~\onlinecite{rom_com,rom}. This fit results in large scattering and unphysical values of the parameters ($R_i<0$, $U<1$~meV).}
\end{figure}
 
Heating even below the lambda point could result in significant effects. Indeed, a small local temperature increase $\delta T\ll T_0$ at the inner cavity surface increases $R_{BCS}$ to:
\begin{equation}
 R_{BCS}(T)=R_{BCS}(T_0)\exp(U\delta T/kT_0^2) 
\label{eq:two}
\end{equation}
Overheating at the inner cavity surface by 50~mK increases $R_{BCS}$ by $\sim 24\%$ at 2.0~K and $U\simeq $1.5~meV for Nb, and by $\sim55\%$ for $\delta T_s = 0.1$~K. Temperature maps of the outer cavity surfaces have routinely revealed local temperature increases,  $\simeq  50-100$~mK at the rf fields $\sim90-100$~mT \cite{Padamsee2009,gc}, which indicates higher $\delta T>\Delta T$ in hotspots at the inner surface \cite{gc}. It is generally difficult to accurately extract $T_s$ from the thermometry data since the temperature of the outer surface cannot be measured directly because only a fraction of the heat is transferred to the thermometers. Moreover, hotspots can result in significant admixture of the BCS component to the averaged residual resistance $\bar{R}_i$ and its significant temperature and field dependencies at strong fields \cite{gc}. This effect masks the intrinsic field and temperature dependencies of $R_s$, so the separation of heating effects is essential.

Romanenko and Grassellino apparently misunderstood our interpretation of the microwave suppression of $R_s$ which was based on the analysis of the experimental data and the well-known results of the BCS theory. The mechanism suggested in Ref. \onlinecite{apl_Qrise} not only explains the observed extended increase of $Q(H)$ but also predicts a logarithmic field dependence of $A(H)$ in excellent agreement with the data. The opinion of Romanenko and Grassellino stems from the model of Ref. \onlinecite{rom} which assumes that the BCS contribution can be unambiguously separated from the residual resistance at any rf field because $R_i$ dominates at $T\ll T_c$. The latter is based on the incorrect postulate that the residual resistance is physically unrelated to the quasiparticle BCS contribution in the simplest version of the BCS model in which the density of states $N(E)$ vanishes at all energies below the gap $|E|<\Delta$, and $R_{BCS}$ at high fields can be evaluated a-priori using the Mattis-Bardeen formula which is only valid in the limit of $H\to 0$.  

The quasiparticle BCS surface resistance can be intertwined with the residual resistance, particularly at high fields. One mechanism related to the vortex hotspots\cite{gc} was already mentioned above. Another mechanism of coupling berween $R_i$ and $R_{BCS}$ is due to the current pairbreaking induced by the rf field which makes $U(H)$ smaller than $\Delta$, and a finite $N(E)$ at $E<\Delta$ due to the sub-gap states at $H=0$ which have been revealed by numerous tunneling experiments on all superconducting materials (see, e.g., Ref.  \onlinecite{ag} and references therein).  Because small but finite $N(E)$ at $E<\Delta$ gives rise to a weakly temperature dependent contribution attributed to the residual resistance, the latter is a natural part of the quasiparticle (BCS) contribution taking into account the realistic $N(E)$ observed on superconducting materials (other extrinsic contributions to $R_i$ were discussed in Ref. \onlinecite{ag}).  As the rf field amplitude increases, the rf currents further broaden $N(E)$ resulting in the microwave suppression of $R_s(H)$. The importance of the sub-gap states and the interplay of the "natural" and the rf broadening of $N(E)$ and their effect on the observed increase of $Q(H)$ with $H$ was discussed at the end of our Letter \cite{apl_Qrise}. 

In conclusion, none of the claims of Romanenko and Grassellino is relevant or backed by scientific arguments. At the same time, the physical conclusions based on the deconvolution of $R_{BCS}$ and $R_i$ using the procedure of Ref. \onlinecite{rom} are questionable because of the incorrect model assumptions, and poor stability and accuracy of the Arrhenius fit for the insufficient number of the temperature datapoints, as discussed above.

\end{document}